\documentclass[aps,prb,twocolumn,superscriptaddress]{revtex4-2}

\usepackage{graphicx}% Include figure files

\begin{document}

%\preprint{}

\title{Mixed-Topology Bubbles Mediate Dipole Skyrmion Formation in Fe/Gd Multilayers}


\author{Jacques A. Reddinger}
\email[Contact author:]{jacquesr@uoregon.edu}
\author{William S. Parker}
\affiliation{Department of Physics, University of Oregon, Eugene, OR 97403, USA}
\author{Sergio A. Montoya}
\author{Eric E. Fullerton}
\affiliation{Center for Memory and Recording Research, University of California, San Diego, La Jolla, CA 92093, USA}
\author{Benjamin J. McMorran}
\affiliation{Department of Physics, University of Oregon, Eugene, OR 97403, USA}


\date{\today}

\begin{abstract}
    Centrosymmetric multilayers provide a route to form dipole-stabilized skyrmions whose topology arises from competition among exchange, anisotropy, and long-range dipolar fields rather than from the competition between exchange and Dzyaloshinskii-Moriya interactions. In Fe/Gd multilayers, skyrmion lattices are commonly formed by collapsing stripe spin textures, but this process produces a mixture of dipole skyrmions and topologically trivial bubbles, indicating that topology selection during collapse is not yet deterministic. While these spin textures exhibit depth-dependent domain walls which smoothly twist from Bloch-like centers to Néel-like surfaces, their integer topological charge does not vary. Here we combine room-temperature Lorentz transmission electron microscopy (LTEM), single-image transport-of-intensity-equation (SITIE) magnetic induction reconstruction, micromagnetic simulations, and domain tracking to identify the topological transition pathways that connect field-driven stripe collapse to final bubble topology. Experimentally, we observe bubble domains whose projected induction patterns are inconsistent with skyrmions or trivial bubbles and are instead consistent with mixed-topology bubbles. Simulations show that stripe collapse initially nucleates trivial bubbles, which subsequently relax either by Bloch-line rotation or through a mixed-topology intermediate containing a topologically trivial Bloch-like interior and skyrmion Néel-like surface layers. Depth-resolved energetic analysis shows that demagnetizing fields favor Néel-cap expansion, while coupling between Bloch lines and the tilted Zeeman field can stabilize the trivial core or reverse the transition. These results support a mechanism in which mixed-topology bubbles mediate dipole skyrmion formation in Fe/Gd multilayers and suggest routes for controlling topology selection in centrosymmetric magnetic multilayers.
\end{abstract}

%\keywords{}

\maketitle

\section{Introduction}

Magnetic bubbles were first reported by Bobeck et al. in centrosymmetric rare-earth iron garnet films with strong perpendicular magnetic anisotropy \cite{bobeckMagneticBubblesEmerging1975}. These bubbles are stabilized by a balance of exchange, anisotropy, and dipolar interactions and can host distinct domain-wall chirality configurations: Type-I bubbles feature a uniform chirality and Type-II bubbles feature two segments of opposite chirality. The discovery of magnetic skyrmions in bulk MnSi renewed interest in small, circular magnetic domains with nontrivial topology \cite{muhlbauerSkyrmionLatticeChiral2009b}. Their topological properties, particle-like dynamics, and proposed use in racetrack-architecture devices have made skyrmions a central target for spintronic materials design \cite{fertSkyrmionsTrack2013e}. While skyrmions are often associated with chiral magnets in which the Dzyaloshinskii-Moriya interaction (DMI) fixes the domain-wall handedness, a related topological structure can also be stabilized in centrosymmetric systems by dipolar interactions \cite{yuMagneticStripesSkyrmions2012b,montoyaTailoringMagneticEnergies2017,chakrabarttyTunableRoomTemperature2022}. Type-I bubbles share the same topological charge as skyrmions found in chiral magnets and are often referred to as dipole skyrmions. Conversely, Type-II bubbles, which have zero topological charge, are often referred to as topologically trivial bubbles. Centrosymmetric magnets therefore allow skyrmions and trivial bubbles to coexist under ambient conditions \cite{desautelsRealizationOrderedMagnetic2019d}.

Fe/Gd multilayers are a model centrosymmetric system capable of hosting stripes, dipole skyrmions, trivial bubbles, antiskyrmions, and bound skyrmion pairs \cite{montoyaTailoringMagneticEnergies2017,leeSynthesizingSkyrmionBound2016d,heiglDipolarstabilizedFirstSecondorder2021b}. In this material family, dipole skyrmions and trivial bubbles can both be stabilized at room temperature, driven by spin-orbit torques, and detected through transport signatures associated with field-dependent stripe and skyrmion phases \cite{montoyaTailoringMagneticEnergies2017,montoyaSpinorbitTorqueInduced2018c,montoyaTransportPropertiesDipole2022c}. These textures form from similar parent stripe textures and can coexist across broad field ranges, yet nominally similar stripe-collapse events may produce different final bubble topologies \cite{titzePathwaysBubbleSkyrmion2025a}. Prior work established Fe/Gd multilayers as a platform for dipole skyrmions and skyrmion lattices, but the intermediate textures that determine topology selection during stripe-collapse remain unexplored. Understanding these intermediate textures is essential for deterministic skyrmion generation.

Here we report LTEM evidence for bubble spin textures in Fe/Gd multilayers whose projected magnetic induction is consistent with mixed, depth-dependent topology. Because SITIE reconstructs projected in-plane induction rather than layer-resolved magnetization, this evidence must be interpreted together with simulations capable of resolving depth-dependent topology. We therefore place the experimental signatures first and use micromagnetic simulations to identify compatible topology and energetic signatures.

We identify a continuous transition pathway from stripe to Type-II (trivial bubble) to Type-I (dipole skyrmion) spin textures mediated by a mixed-topology Type-III spin texture, with depth-localized topology changes at partial-Bloch-line termini. Simulations reveal two relaxation pathways for frustrated Type-II bubbles formed from the collapse of stripes under an anti-aligned in-plane Zeeman field: Bloch-line rotation or a topological transition through a metastable Type-III spin texture. The competition between local demagnetizing fields and Bloch-line coupling to the in-plane Zeeman field determines whether a Type-III bubble becomes a Type-I bubble or reverts toward Type-II bubble.

\section{Background}

In magnetic materials, out-of-plane (perpendicular) magnetic domains are usefully described by their domain walls, where the magnetization reverses between neighboring domains. The domain-wall helicity is the orientation of the in-plane magnetization relative to the domain-wall normal. In multilayer films where perpendicular domains arise from a thickness driven spin reorientation, this helicity is depth dependent. Domain walls are approximately Bloch-like near the midplane and acquire Néel caps at the top and bottom surfaces to reduce magnetostatic energy \cite{hubertMagneticDomains1998, lemeshTwistedDomainWalls2018b, parkerHybridSkyrmionsMagnetic2024c}. Neutron scattering measurements and multimodal electron microscopy of Fe/Gd multilayers have directly confirmed that dipolar skyrmions in this system have hybrid three-dimensional structure with depth-dependent domain-wall character \cite{liyanageThreedimensionalStructureHybrid2023b, parkerRealSpaceImaging2025}. In the absence of appreciable DMI, the clockwise and counterclockwise Bloch components of the domain walls are nearly degenerate, so helicity can be selected dynamically by local fields and defects.

\begin{figure*}
\includegraphics{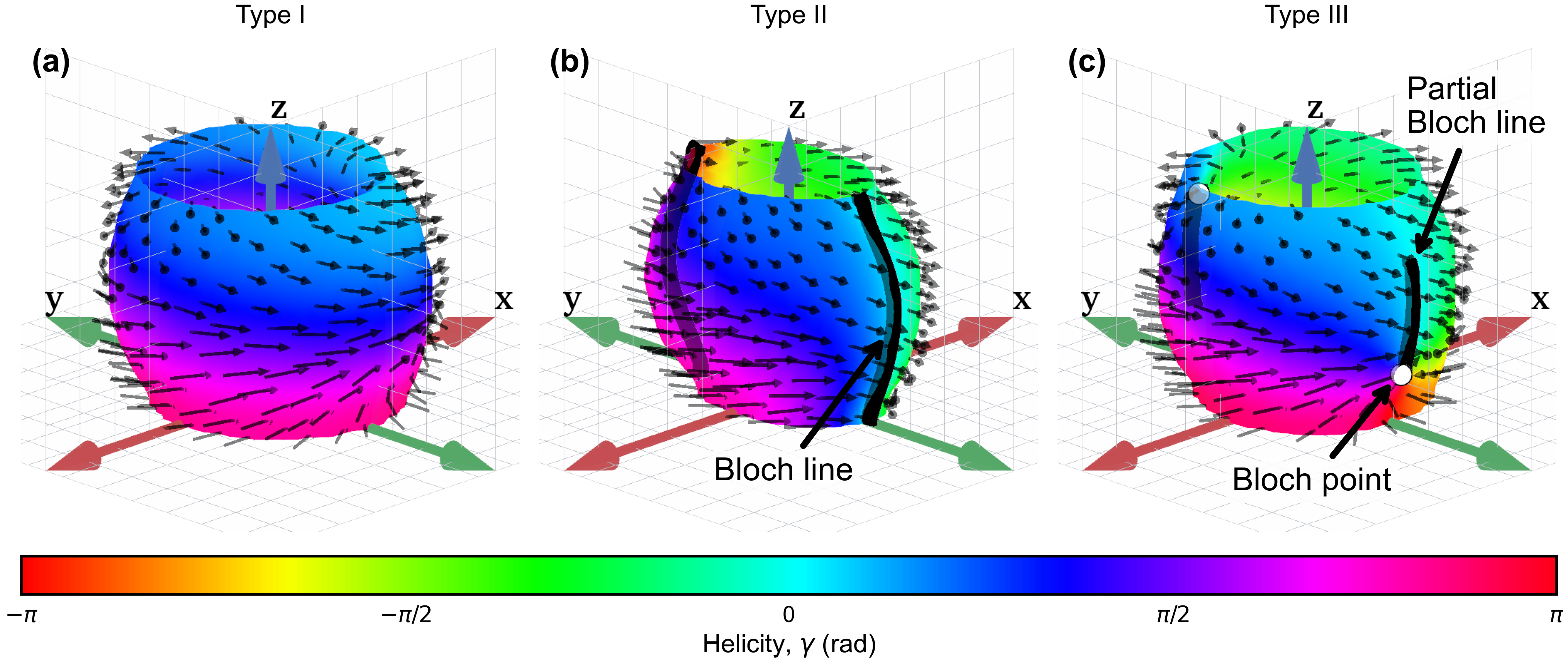}
\caption{\label{fig:fig1} Simulated topology and domain-wall structure of different bubble types in centrosymmetric Fe/Gd multilayers. (a) Type-I bubble (dipole skyrmion) domain wall shell with constant in-plane winding through the film thickness. (b) Type-II bubble (trivial bubble) containing a pair of Bloch lines (black lines) on either end which extend through the entire film thickness and separate opposite chirality regions. (c) Type-III bubble (mixed topology bubble) with topologically trivial core layers containing a pair of partial Bloch lines which terminate in the film and topologically non-trivial surface layers. One terminus of each partial Bloch lines carries a Bloch point (white circle) where the magnetization forms a singularity. Black-colored arrows indicate the in-plane magnetization at the domain wall. Domain wall surface color indicates the helicity, the in-plane magnetization relative to the surface normal.}
\end{figure*}

A Bloch line is a region along the domain wall where the Bloch-wall chirality reverses. Type-I bubbles [Fig. \ref{fig:fig1}(a)] have domain-wall magnetization that winds once around the bubble without Bloch lines and therefore carry nonzero skyrmion number in each layer. Type-II bubbles [Fig. \ref{fig:fig1}(b)] contain two Bloch lines and have zero net winding \cite{yuMagneticStripesSkyrmions2012b}. Both Type-I and Type-II bubbles can possess depth-dependent Néel caps, but their two-dimensional topology remains the same through the film thickness: Type-I remains skyrmion-like in every layer, whereas Type-II remains topologically trivial in every layer.

We reserve the term Type-III [Fig. \ref{fig:fig1}(c)] for a more general bubble with mixed topology through the film thickness. In this spin texture, interior layers are topologically trivial and feature two Bloch lines where the chirality reverses, while one or both surfaces have uniform chirality and thus are not topologically trivial. The transition between these regions is bounded by partial Bloch lines whose termini form Bloch points, singularities in the magnetization. Thus, an ordinary Type-II bubble with flux-closure Néel caps is not automatically Type-III; the defining feature is a layer-resolved change in winding/topological character. Type-III bubbles have been predicted in simulation and observed in other materials \cite{wangFormationMagneticBiskyrmions2022,kongObservationHybridMagnetic2023}, and stripe-to-skyrmion formation pathways have been studied in chiral multilayers \cite{chenThermalEvolutionSkyrmion2022}, but the role of mixed-topology intermediates during stripe collapse into Type-I bubbles in Fe/Gd multilayers has not been established.

LTEM is sensitive to the projected in-plane induction of the film. It encodes induction patterns associated with Bloch-line structure and domain-wall winding, but it does not by itself reconstruct the full three-dimensional magnetization. The projected Type-III bubble structure appears similar to the projected Type-II bubble in LTEM. This ambiguity motivates the combined experimental-simulation approach used here. The experimental identification of mixed-topology bubbles in this work is therefore cautious: SITIE identifies projected signatures consistent with a Type-III bubble, and micromagnetic simulations provide the depth-resolved interpretation.

\section{Methods}

\subsection{Experimental Methods}

Room-temperature LTEM measurements were performed using an FEI ThemIS at the National Center for Electron Microscopy. The sample was a Ta(5 nm)/[Fe(0.34 nm)/Gd(0.41 nm)]$_{120}$/Ta(5 nm) multilayer film sputter deposited on a 50 nm SiN TEM membrane at room temperature under an ultrahigh vacuum environment at 3 mTorr Ar. The sample was saturated by exciting the objective lens to +20\% corresponding to approximately 6 kOe at the sample. Phase-contrast Fresnel images were acquired with a defocus of 500 µm while sweeping the objective-lens from +5\% to -5\% in 0.1\% steps. Stripe collapse began around 3.5\% excitation, corresponding to $|\mathbf{H}|=1.05\pm0.05$ kOe.

The sample holder was tilted by $4^\circ$ from the electron-beam direction to introduce an in-plane field component of $|\mathbf{H}|=73\pm4$ Oe and frustrate parallel stripe domains. The field sweep drove transitions from positive saturation, through stripes, and into negative-field bubbles. Fresnel images were processed with a bandpass filter to reduce shot noise and background. Magnetic induction vector fields were reconstructed from the processed Fresnel images using the single-image transport-of-intensity equation, following prior LTEM/SITIE approaches for skyrmionic materials and domain-wall chirality mapping \cite{chessDeterminationDomainWall2017d,chessStreamlinedApproachMapping2017d}.

Regions of interest containing bubbles with apparent trivial topology (Type-II or Type-III) were extracted from the magnetic induction images using a multi-scale ridge-detection algorithm \cite{lin_lxfhfutridge-detector_2026, stegerUnbiasedDetectorCurvilinear1998}. Only Type-II or Type-III bubbles that could be traced to a stripe at, or immediately adjacent to, the same position before nucleation were included in the analyzed population. Gaps between detected ridge segments were closed by projecting from the segment endpoints. Within this eligible population, bubble categorization was performed manually using a single criterion. Bubbles without a discernible core induction stream between the two Bloch lines were classified as Type-II [Fig. \ref{fig:fig1}(b)], whereas those with the stream were classified as candidate Type-III [Fig. \ref{fig:fig1}(c)]. Area, eccentricity, helicity profile, or magnitude of projected magnetic induction were not considered during classification.

Some candidate Type-III bubbles had weak contrast relative to the surrounding magnetic texture, which is consistent with a mixed-depth wall: LTEM phase contrast is governed by the projected in-plane induction, and replacing part of a through-thickness Bloch-like wall with skyrmion-like surface winding reduces the effective Bloch-wall contribution to the projection \cite{chessDeterminationDomainWall2017d,chessStreamlinedApproachMapping2017d,loudonImagesBiskyrmionsShow2019,pollardObservationStableNeel2017b}. Accordingly, the experimental classifications are used as projected signatures and are interpreted through comparison with simulated Type-I, Type-II, and Type-III bubble induction maps rather than as standalone measurements of layer-resolved topology.

\subsection{Statistical Analysis}

Statistical comparisons were performed after manual classification, so projected area, eccentricity, helicity profile, and projected-induction magnitude were not used to identify Type-II and Type-III bubbles. For bubble-level comparisons, each unique experimental bubble was treated as one observation. Projected area, eccentricity, median radial distance, and median projected-induction magnitude were compared between Type-II and candidate Type-III bubbles using two-sided Mann-Whitney U tests \cite{mannTestWhetherOne1947}. These results were further refined to q, the Benjamini-Hochberg corrected p values, where smaller q values indicate stronger evidence that two populations are distinct after accounting for the four simultaneous property comparisons \cite{benjaminiControllingFalseDiscovery1995a}. To test whether the populations were separated jointly across all measured properties, projected area, eccentricity, radial distance, and projected magnetic induction magnitude were standardized and compared with an energy-distance permutation test using 10,000 permutations, following procedures outlined in \cite{szekelyEnergyStatisticsClass2013,andersonNewMethodNonparametric2001a}.

\subsection{Computational Methods}

Micromagnetic simulations were performed using MuMax3, which numerically solves the Landau-Lifshitz-Gilbert equation \cite{vansteenkisteDesignVerificationMuMax32014}. Material parameters were taken from Montoya et al. \cite{montoyaTailoringMagneticEnergies2017} and are representative of Fe/Gd multilayer thin films.

The simulated geometry was a 2.048 µm × 2.048 µm × 80 nm slab of (3.2 nm, 3.2 nm, 3.2 nm) voxels with periodic boundary conditions in the film plane and random initial magnetization. Stripe collapse was induced by sweeping an external magnetic field from 330 mT to -300 mT, tilted $4^\circ$ from the film normal. To prepare the stripes, the field was swept from 330 mT to -200 mT in 5 mT increments using energy minimization, followed by 20 ns of LLG evolution at -200 mT. From this state, a time-dependent field sweep from -200 mT to -300 mT was applied over 100 ns. Damping was set to $\alpha=1$ to approximate the fast dynamics compared to the slowly varying external field. Magnetization and energy components were recorded every 0.25 ns. Simulations were carried out at both 0 K and 273 K (see supplementary materials). Both runs showed similar behavior. The 0 K run was analyzed with the domain-tracking algorithm to minimize pre-processing.

A custom python package, \verb+Domain Tracker+ \cite{reddingerJareddiDomaintracker2026}, processed the MuMax3 output and extracted structural and energetic information for each domain. Contours from individual z slices were stitched into three-dimensional \verb+Domain+ objects, linked across time by center-of-mass proximity and intersection-over-union metrics, and classified as stripes or Type-I, Type-II, or Type-III bubbles. The resulting stripe-to-bubble histories enabled event counting, topology tracking, and depth-resolved energy analysis. The \verb+Domain Tracker+ package and analysis scripts used in this work are publicly available on GitHub.

These data provide the temporal and energetic resolution needed to interpret experimentally observed projected induction signatures and to resolve topology changes that are inaccessible to LTEM/SITIE alone.

\section{Results}

\subsection{Experimental signatures of mixed-topology bubbles}

\begin{figure*}
\includegraphics{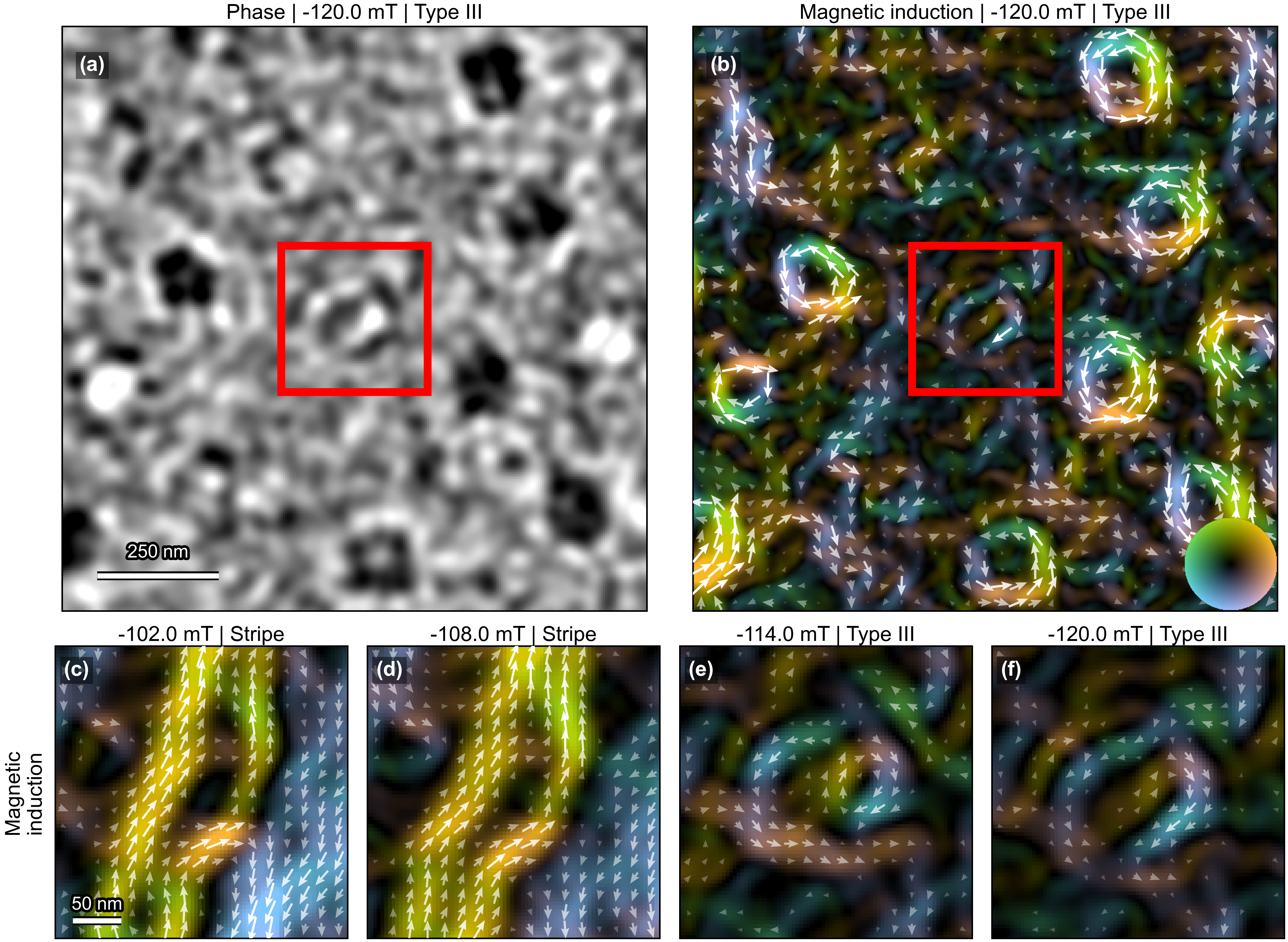}
\caption{\label{fig:fig2}
Experimental LTEM/SITIE evidence for mixed-topology bubbles after stripe collapse. (a) Reconstructed phase image and (b) SITIE-reconstructed projected magnetic induction map of magnetic spin textures after collapse into cylindrical spin textures; the red box marks a mixed-topology bubble surrounded by Type-I and Type-II bubbles. (c)-(f) SITIE induction image series of the boxed region during an objective-lens sweep from -102.0 mT to -120.0 mT. The newly formed Type-III bubble develops a skyrmion-like morphology while preserving a Type-II-like projected induction signature, consistent with a mixed-topology bubble.
}
\end{figure*}

LTEM Fresnel images acquired during the tilted field sweep show stripe collapse followed by the appearance of isolated bubble domains with different topologies [Fig. \ref{fig:fig2}(a)]. SITIE reconstruction of the projected induction [Fig. \ref{fig:fig2}(b)] reveals that most bubbles are consistent with either a topologically non-trivial, cylindrical (Type-I bubble) texture or topologically trivial, cylindrical (Type-II bubble) texture. Within the stripe-derived topologically trivial population, however, a subset contains a discernible core induction stream connecting the two Bloch lines. This core stream is the selected criterion used to assign candidate Type-III bubbles. One such domain is marked in Fig. \ref{fig:fig2}(b), and its evolution during stripe collapse is shown in Figs. \ref{fig:fig2}(c)-\ref{fig:fig2}(f). The image montage shows a stripe domain anti-aligned to the in-plane Zeeman field collapsing into an atypical bubble with two joined vertices, resembling the bound-pair morphology previously reported in Fe/Gd multilayers \cite{leeSynthesizingSkyrmionBound2016d} and the projected biskyrmion-like LTEM contrast analyzed by Loudon et al. \cite{loudonImagesBiskyrmionsShow2019}. The domain wall also exhibits a relatively weak projected induction magnitude compared with neighboring bubbles.

The key experimental observation is therefore not a direct measurement of three-dimensional topological charge, but the appearance of projected induction patterns that cannot be explained cleanly by a purely Type-I or Type-II bubble that is projected across the thickness. This motivates comparison with simulated projected induction from Type-III bubbles, in which Type-I-bubble surface layers coexist with a Type-II-bubble interior.

\subsection{Projected-induction comparison with simulations}

\begin{figure}
\includegraphics{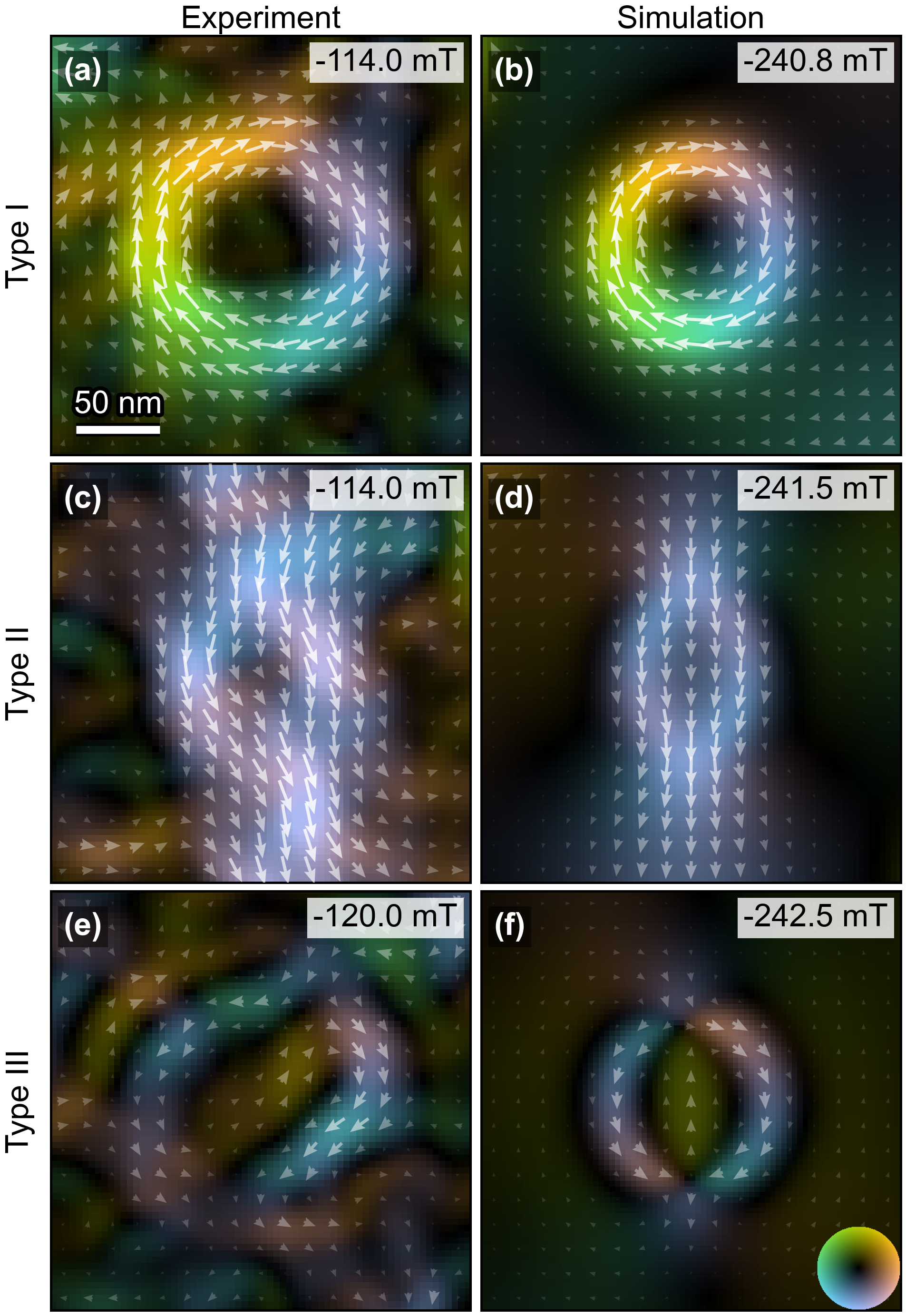}
\caption{\label{fig:fig3}
Experimental (left) and simulation (right) comparison of projected induction signatures of different topology spin textures: (a, b) Type-I, (c, d) Type-II, and (e, f) Type-III bubbles. The color-wheel and white-colored arrows show the magnitude and direction of the in-plane magnetic induction. This enables a comparison of the chirality and morphology between experimental and simulated bubble spin textures.
}
\end{figure}

In Fig. \ref{fig:fig3}, projected magnetic induction maps of Type-I, Type-II, and Type-III bubbles are compared between experiment and simulation. The Type-I bubbles show continuous circulation around the domain wall, while the Type-II bubbles show the characteristic asymmetry associated with a Bloch-line pair. Type-III bubbles were identified by a discernible core induction stream connecting two Bloch lines, a feature reproduced by simulated Type-III bubble projections. This projected signature does not by itself determine the three-dimensional topology, but it provides the experimental criterion used for comparison with the depth-resolved simulations.

We next asked whether the core-stream-defined populations differ in projected morphology and induction structure. After classification, we compared Type-II and Type-III bubbles using domain-level morphology, pixel-level induction statistics, and binned helicity profiles. These quantities were not used to assign bubble type; they provide independent post-classification checks on whether the two experimental populations remain distinct.

\begin{figure*}
\includegraphics{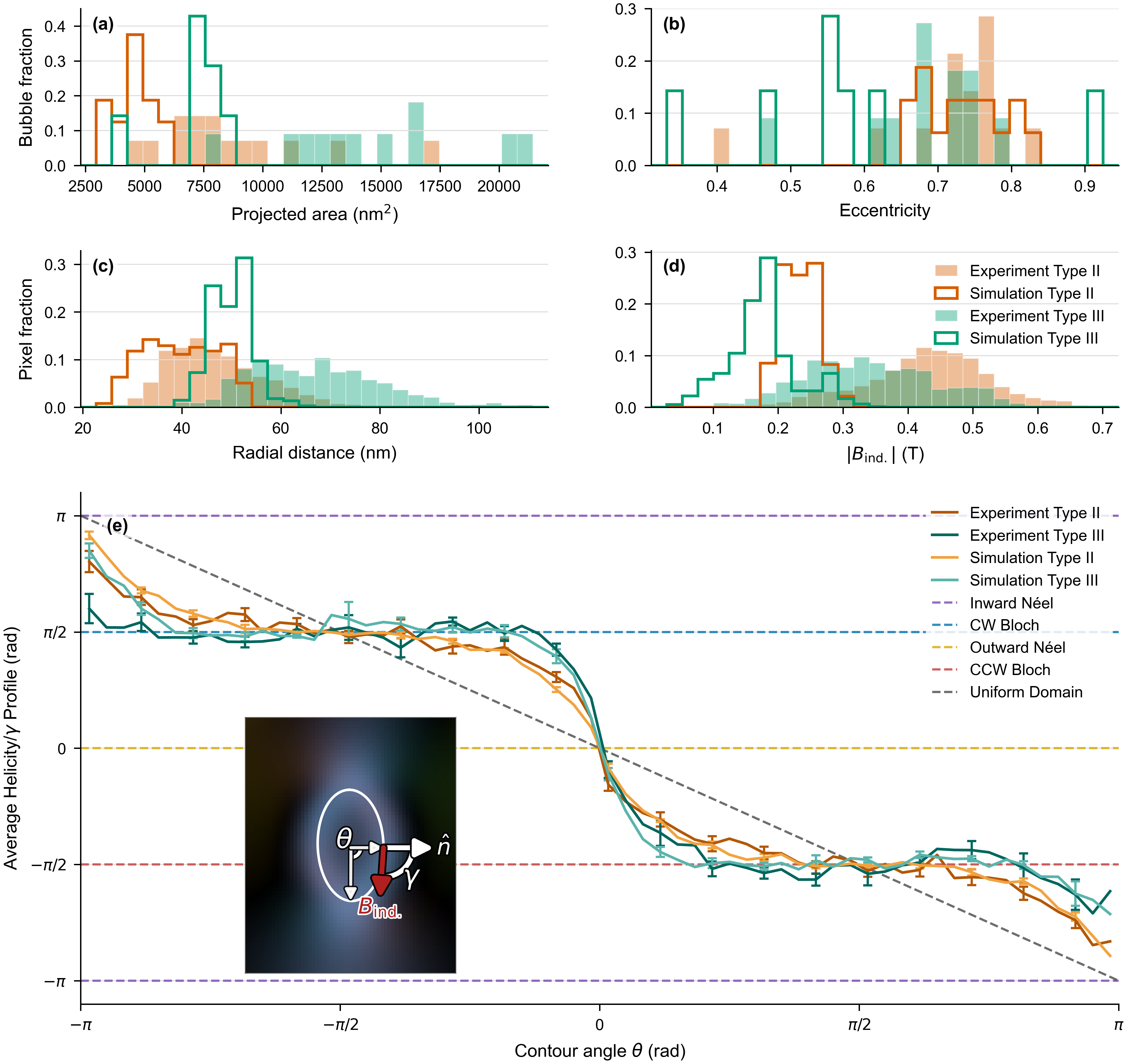}
\caption{\label{fig:fig4}
Experimental and simulation property comparison of Type-II and Type-III bubbles. Bubble-level distributions of (a) projected area and (b) eccentricity compare the morphological properties of the two populations. Ridge-point-level distributions of (c) radial position and (d) projected magnetic induction magnitude characterize the internal induction structure. (e) Domain-wall helicity as a function of contour angle for Type-II and Type-III bubbles. The inset defines contour angle as the angle relative to the $-\hat{y}$ direction and helicity as the angle between the outward domain-wall normal and the local projected magnetic induction. The helicity profile is calculated per bubble by averaging helicity into contour angle bins of width $\pi/30$. These per bubble helicity profiles are then averaged by type.
}
\end{figure*}

Aggregating over many Type-II and III bubbles makes these trends clearer. After filtering, the domain-level histograms contain 14 experimental Type-II and 11 experimental Type-III bubbles, together with 16 simulated Type-II and 7 simulated Type-III bubbles. The pixel- or ridge-point-level histograms contain 1486 experimental Type-II and 1657 experimental Type-III points, and 1959 simulated Type-II and 945 simulated Type-III points. Ridge-point-level histograms are shown descriptively to visualize the internal distributions within the classified bubble populations. Fig. \ref{fig:fig4} shows histograms of ridge and induction properties from experiment and simulation. Although the absolute values differ between experiment and simulation, the relative trends agree: Type-III bubbles are generally larger, less eccentric, and have weaker projected induction than Type-II bubbles. The weaker projected induction is physically consistent with a reduced effective Bloch-wall depth in Type-III bubbles, because LTEM/SITIE measures the through-thickness projected in-plane induction rather than the full three-dimensional magnetization \cite{chessDeterminationDomainWall2017d,chessStreamlinedApproachMapping2017d,loudonImagesBiskyrmionsShow2019,pollardObservationStableNeel2017b}.

Treating each unique experimental bubble as one statistical observation, Type-II and Type-III bubbles form distinct populations. Type-II bubbles had a median projected area of $7.92 \times 10^3$ nm$^2$, median radial distance of 45.5 nm, median eccentricity of 0.736, and a median projected magnetic induction of 0.43 T. Type-III bubbles had a median projected area of $1.40 \times 10^4$ nm$^2$, median radial distance of 63.7 nm, median eccentricity of 0.678, and a median projected magnetic induction of 0.34 T. Using the statistical methods, the two populations were compared over these four properties. Below, q denotes the Benjamini–Hochberg adjusted p-value, with smaller values providing stronger statistical evidence that Type-II and Type-III bubbles have distinct distributions of the measured property, after accounting for multiple comparisons. The results confirmed the two populations differ significantly in projected area ($q = 0.003$), radial distance ($q = 3.62 \times 10^{-4}$), and projected magnetic induction ($q = 0.003$). While there was a difference in eccentricity, it was not significant (q = 0.163). The joint distributions of projected area, eccentricity, radial distance, and magnetic-induction magnitude also differed significantly between Type-II and Type-III bubbles (energy-distance permutation test on standardized features, $E = 2.36$, $p_\textrm{perm} = 1.0 \times 10^{-4}$; 10,000 permutations) \cite{szekelyEnergyStatisticsClass2013, andersonNewMethodNonparametric2001a}. The domain-wall helicity profiles also differ. The Type-III helicity profile changes more sharply near the Bloch-line position ($\theta=0$) and remains flatter away from it, like Type-I bubbles. Alternatively, Type-II bubbles are closer to a linearly decreasing helicity, like a uniform magnetization.

These post-classification differences characterize the two experimental populations; the precursor-stripe evolution and comparison with simulated projections provide the model-informed basis for interpreting the core-stream signature as consistent with Type-III topology. With the experimental Type-III bubbles defined by their projected induction signatures, we next ask how they form during stripe collapse. The simulations provide the time resolution needed to connect the observed post-collapse textures to the underlying topology-changing pathways.

\subsection{Simulated stripe collapse nucleates frustrated Type-II bubbles}

\begin{figure*}
\includegraphics{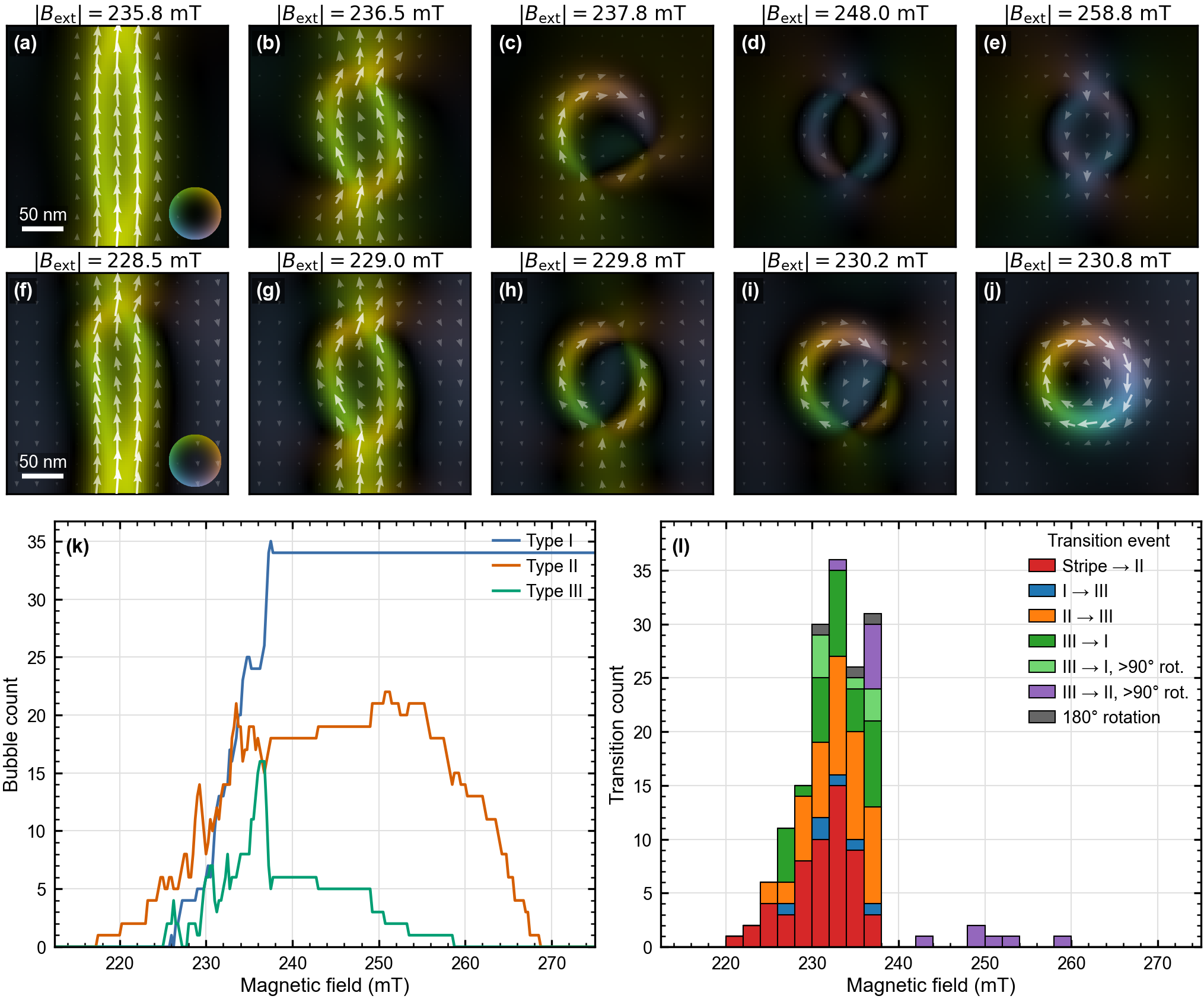}
\caption{\label{fig:fig5}
Simulated stripe collapse and transitions statistics under a perpendicular applied field sweep. The simulated geometry was a 2.048 µm × 2.048 µm × 80 nm slab of (3.2 nm, 3.2 nm, 3.2 nm) voxels with periodic boundary conditions in the film plane. (a)-(j) Simulated projected magnetic induction for two representative stripe-collapse trajectories. In both cases, stripe-domain pinch-off nucleates a Type-II bubble that passes through a Type-III state; the upper sequence relaxes back toward Type-II topology after Bloch-line rotation, whereas the lower sequence converts to a Type-I bubble. Field values label each frame. (k) Bubble populations versus applied field from one simulation run show that stripe collapse initially produces Type-II bubbles before Type-III and Type-I states appear. (l) Transition counts between bubble types from one simulation run show no direct Type-I/Type-II conversion, indicating that Type-III bubbles mediate topological transitions in the simulated ensemble.
}
\end{figure*}

In Fig. \ref{fig:fig5}, representative simulated domains are tracked after stripe collapse. In the simulations, stripes shrink gradually under increasing out-of-plane field until a critical field causes individual stripes to pinch-off. Bloch-line pairs form naturally at the pinch-off points, so the immediate products of stripe collapse are Type-II bubbles that preserve the topology and Bloch-line orientation of the original stripe. These nucleated Type-II bubbles inherit a frustrated orientation relative to the in-plane component of the tilted Zeeman field.

The projected-induction sequences in Figs. \ref{fig:fig5}(a)-\ref{fig:fig5}(j) show two representative outcomes for frustrated Type-II bubbles. In both cases, a stripe pinched off into a Type-II bubble and then accessed a Type-III topology. In one trajectory the Bloch-line pair rotated toward alignment with the tilted field and the bubble returns to Type-II topology; in the other, the surface layers collapsed and the bubble converted to a Type-I bubble.

The ensemble statistics in Figs. \ref{fig:fig5}(k) and \ref{fig:fig5}(l) place these representative trajectories in context. Across 48 tracked post-collapse bubbles and 96 classified transition events, Type-II bubbles appear immediately after stripe collapse and are followed by the emergence of Type-III and Type-I bubbles. No direct Type-II to Type-I transitions (or vice versa) were observed within the simulated ensemble; topological conversion proceeded through Type-III bubbles.

This sequence is consistent with the experimental observation that anti-aligned stripe collapse produces a mixture of Type-I bubbles, field-aligned Type-II bubbles, and Type-III bubbles. The simulations suggest that the frustrated Bloch-line geometry created at pinch-off is resolved by two competing responses: rotation of the Bloch-line pair or topological transition through a Type-III intermediate.

\subsection{Type-III bubbles mediate topology change}

\begin{figure*}
\includegraphics{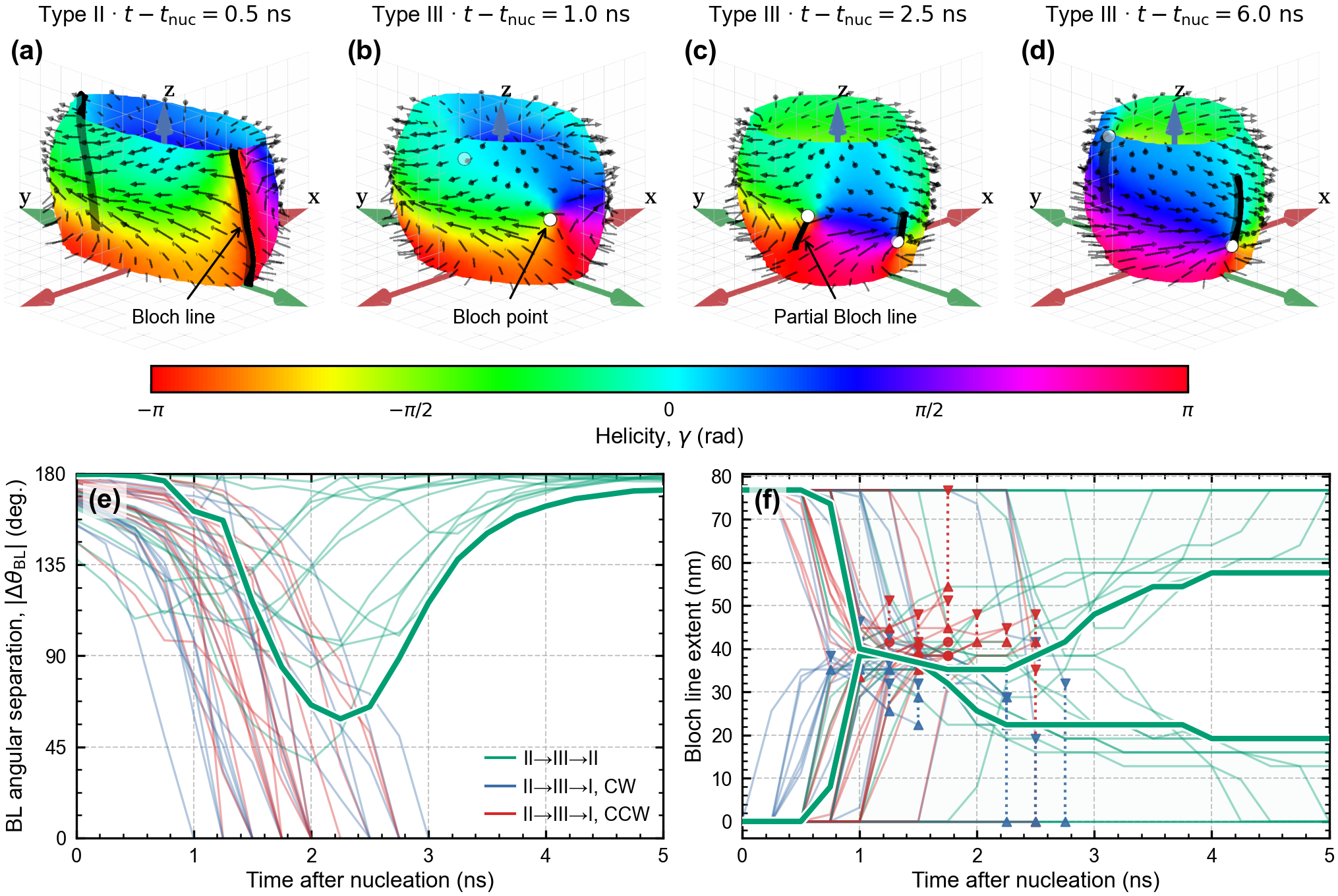}
\caption{\label{fig:fig6}
Depth-resolved topology and Bloch-line dynamics during simulated Type-II to Type-III to Type-I relaxation. (a)-(d) Three-dimensional domain-wall renderings of a tracked bubble with Bloch lines marked by the black lines and Bloch-points marked by dots. After beginning as a Bloch-line-bearing Type-II state, the top and bottom surfaces collapse, reducing the Bloch lines to single-layer points. These points rotate about the bubble core to align with the in-plane component of the external magnetic field, after which the Bloch lines partially expand. (e) Bloch-line orientation versus time after nucleation, separating simple Type-II rotation from topology-changing relaxation. (f) Bloch-line extent versus time after nucleation, showing partial collapse during Type-III formation. Triangles and dashed lines indicate the final Bloch-line extent at transition to Type-I; circles indicate cases in which the final Bloch-line extent at transition was a single layer. The highlighted green lines in (e) and (f) correspond to the bubble evolution displayed in (a)-(d).
}
\end{figure*}

Figure \ref{fig:fig6} resolves the Bloch-line dynamics underlying those transition counts. All Type-I-bubble-forming events pass through Type-III bubbles, whereas bubbles that remain Type-II either rotate directly or briefly enter Type-III before returning to Type-II topology. The distinction is therefore not whether a Bloch-line pair forms at nucleation, but whether that pair rotates apart or collapses through the surface layers.

Figs. \ref{fig:fig6}(a)-\ref{fig:fig6}(d) show the three-dimensional Bloch-line structure of a tracked domain that undergoes a 180° rotation. The Bloch-line extent shrinks to a single layer before recovering to approximately half the sample thickness, while the two Bloch lines rotate around the bubble core with different timing. This trajectory approaches, but avoids, the merging event that would annihilate the Bloch lines and complete a Type-I transition. Fig. \ref{fig:fig6}(e) shows the angular displacement between the two Bloch lines: in Type-I-bubble-forming trajectories the Bloch lines rapidly merge and annihilate, whereas in trajectories that remain Type-II or Type-III this merging event is avoided. Fig. \ref{fig:fig6}(f) shows the Bloch-line extent after nucleation. Every domain that converts to Type-I undergoes collapse of the surface layers into Néel-closure domains.

These dynamics identify the local competition that controls the relaxation pathway: Bloch-line reorientation preserves a Type-II-like core, whereas Néel-closure-domain expansion drives the system toward mixed topology and, in some cases, Type-I bubble formation.

\subsection{Local depth-dependent energetics select the final topology}

\begin{figure}
\includegraphics{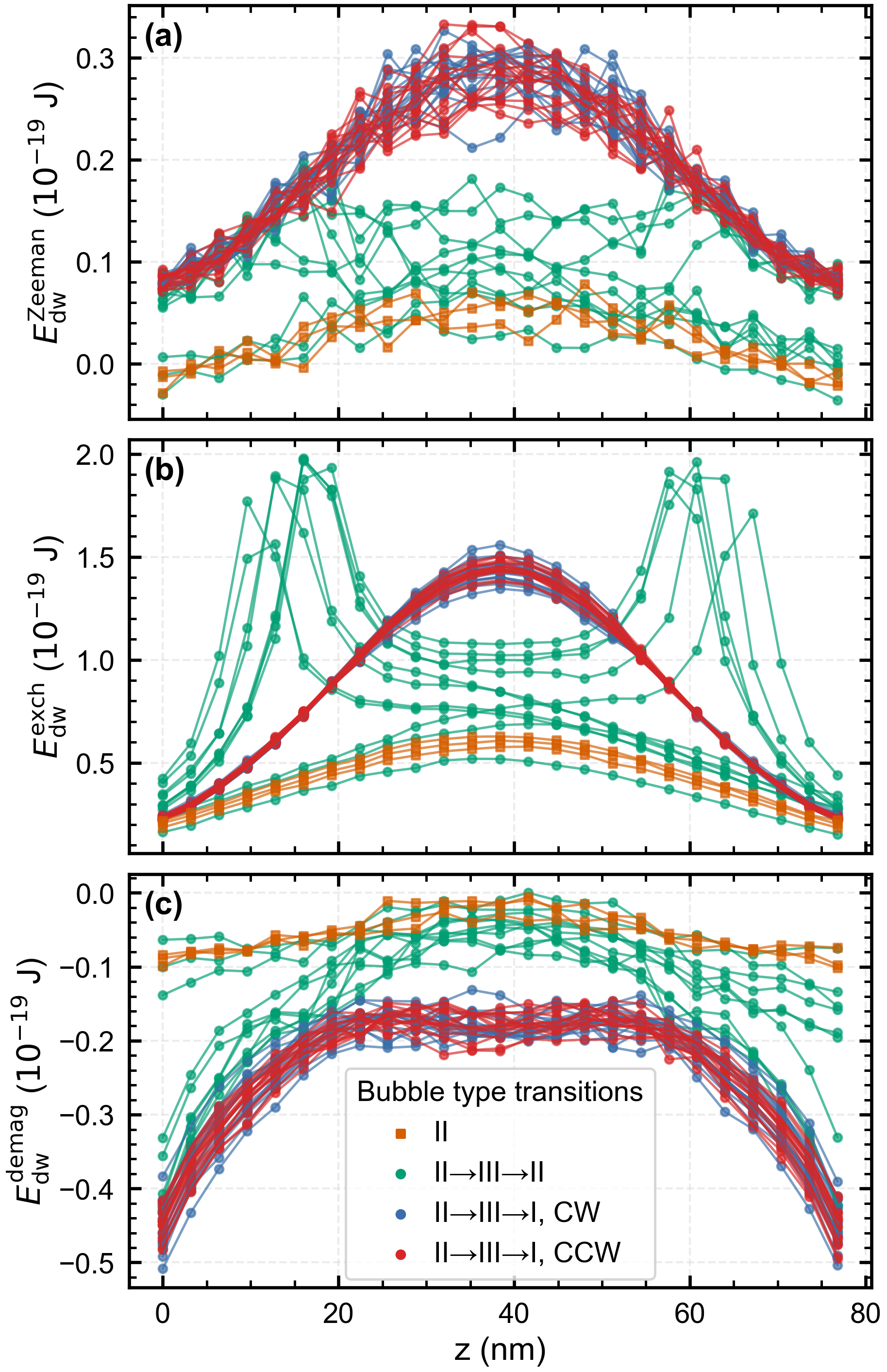}
\caption{\label{fig:fig7}
Depth-dependent energetics and topology of simulated bubble domain walls after nucleation. Layer-resolved differences in (a) Zeeman, (b) exchange, and (c) demagnetizing energy are shown for bubbles grouped by relaxation pathway: Type-II bubbles that remain Type-II, Type-II bubbles that become Type-III and return to Type-II, and Type-II bubbles that become Type-III and continue to Type-I with either final chirality. Type-III-forming pathways occupy an energetic middle ground: skyrmion-like surface layers reduce demagnetizing energy, while the Bloch-line-bearing interior retains favorable coupling to the tilted field.
}
\end{figure}

Depth-resolved energy profiles in Fig. \ref{fig:fig7} probe the energetic competition among exchange, Zeeman, and demagnetizing contributions for bubbles grouped by relaxation pathway. Each bubble's domain-wall energy profile is evaluated 4 ns after nucleation. This snapshot does not by itself give a transition barrier; instead, it provides a layer-resolved comparison of the energetic tendencies that favor persistent Type-II topology, transient Type-III topology, or conversion to Type-I topology after relaxation begins.

Fig. \ref{fig:fig7}(a) shows the Zeeman contribution. Bubbles that retain or recover Type-II-like Bloch-line structure have reduced Zeeman energy where the Bloch lines align favorably with the in-plane component of the Zeeman field. Bubbles that proceed toward Type-I topology lose this favorable Bloch-line contribution as the surface layers collapse and the Bloch lines or Bloch points annihilate.

Fig. \ref{fig:fig7}(b) shows the exchange contribution. Type-II bubble domain walls have reduced exchange energy near their Bloch lines, where the domain-wall curvature is reduced. Pathways that access Type-III bubbles exhibit exchange-energy peaks near the film surfaces, consistent with the rapid magnetization variation at the ends of partial Bloch lines and near Bloch-point-like singularities. Type-I-bubble pathways vary more smoothly once the Bloch lines collapse, but they retain higher exchange energy in the Bloch-like core where the domain-wall width is smaller.

Fig. \ref{fig:fig7}(c) shows the demagnetizing contribution. Pathways that access Type-III or Type-I topology lower the demagnetizing energy in the surface layers by forming Néel caps and improving flux closure. Bubbles that remain Type-II retain less favorable surface demagnetizing energy, while Type-II-like and Type-III-like interiors both carry a demagnetizing-energy cost associated with the Bloch-line-bearing core.

Together, these trends explain why the Type-III bubbles can form during the relaxation process. Local demagnetizing fields favor Néel-cap expansion at the surfaces, while exchange penalizes the Bloch points and partial Bloch lines required to connect skyrmion-like surface layers to a Type-II-like interior. The mixed topology lowers magnetostatic energy at the surfaces while preserving Bloch-line structure in the core. Bubbles that do not access the Type-III pathway instead rotate, lowering the Zeeman energy associated with the initially unfavorable Bloch-line orientation. Thus, the energetics connect the projected Type-III-like signatures observed experimentally to the topology-changing pathways resolved in simulation.

The final topology depends on the competition between Néel-cap expansion and Bloch-line stabilization which itself depends on the local energetic environment where the spin texture is forming. If the demagnetizing-energy gain dominates before the Bloch lines reorient, the Bloch lines collapse and the bubble becomes a Type-I dipole skyrmion. If Bloch-line rotation occurs first, Zeeman coupling stabilizes the Type-II-like core, and the bubble can revert toward Type-II topology.

\section{Discussion}

These results clarify the role of mixed-topology bubbles in Fe/Gd multilayers. LTEM/SITIE provides projected induction evidence for bubbles that are not well described as purely Type-I or Type-II bubbles. Simulations show that such projected signatures are naturally produced by Type-III bubbles, whose surfaces and core carry different two-dimensional topology.

This interpretation also explains why topology selection during stripe collapse appears stochastic in imaging experiments. The relevant control parameters are local and depth dependent: neighboring domains alter the demagnetizing field, while the tilted external field couples to Bloch-line orientation. Small differences in the local environment can therefore determine whether a nucleated Type-II bubble rotates, becomes a Type-I bubble through the Type-III pathway, or returns to a Type-II bubble.

Quantitative offsets between experiment and simulation likely arise from finite temperature, uncertainty in material parameters, disorder and pinning, and uncertainty in the objective-lens field calibration. These factors shift the critical Zeeman field for stripe collapse: experimental stripe collapse occurs around 110 mT, while simulated stripe collapse occurs around 230 mT. Because bubble size decreases with increasing out-of-plane field, a higher effective collapse field would produce smaller projected domains without changing the topological distinction between Type-II and Type-III populations.

The biskyrmion/Type-II-bubble debate provides useful context for this interpretation. Loudon et al. showed that counter-rotating projected induction vortices can arise from Type-II bubbles viewed at an angle and that the projected B field need not trace the underlying magnetization \cite{loudonImagesBiskyrmionsShow2019}. Our results build on that point by separating the projected signature from the depth-resolved topology and from the formation pathway. A Type-III bubble can retain a Type-II-like projected induction component while containing skyrmion-like surface winding and a Type-II-like interior. Thus, projected LTEM/SITIE contrast alone can obscure the distinction between conventional Type-II bubbles and Type-III bubbles, but the ambiguity can be reduced by combining projection comparisons, population statistics, and layer-resolved simulations.

The experimental evidence must be stated cautiously because projected induction does not uniquely determine the full three-dimensional magnetization. Nevertheless, the agreement among SITIE signatures, statistical ridge properties, simulated Type-III projections, and the energetic pathway resolved in simulation supports mixed-topology bubbles as physical intermediates rather than artifacts of a particular simulation protocol.

\section{Conclusion}

We have carefully explored the stripe-to-skyrmion transition in Fe/Gd multilayers using experimental and computational methods. The results have revealed the formation of an intermediate topological spin texture which we refer to as a Type-III bubble. These Type-III bubbles convert from newly formed achiral Type-II bubbles during stripe collapse due to local demagnetizing fields. During this transition, the achiral surface of the Type-II bubble collapses into a chiral surface, lowering the magnetostatic energy. This process is energetically costly as it generates Bloch points (magnetic singularities) at the boundaries between the topologically distinct chiral surfaces and achiral Bloch cores. The Bloch points, and therefore the boundaries between these regions, are not fixed and the depth-dependent topological profile can vary depending on local energetics. This topological profile is established by the competition between the demagnetizing fields, the exchange cost of the Bloch points, and the coupling between the Zeeman-field and the achiral Bloch cores. This competition also determines whether the metastable Type-III bubble continues this topological transition into a Type-I bubble (dipole skyrmion) or reverts to a Type-II (topologically trivial) bubble. These results suggest routes for deterministic topology control in centrosymmetric magnetic multilayers. Future work will investigate the role of neighboring domains on the final topology.

\section{Acknowledgements}
LTEM data were obtained at the National Center for Electron Microscopy at the Molecular Foundry at Lawrence Berkeley National Laboratory. Work at the Molecular Foundry was supported by the Office of Science, Office of Basic Energy Sciences, of the U.S. Department of Energy under Contract No. DE-AC02-05CH11231, under user proposal No. MFP-10439, NSF grant No. 2105400. SAM acknowledges support from the U.S. D.O.D. The authors acknowledge Research Advanced Computing Services (RACS) at the University of Oregon for providing computing resources that contributed to the results reported in this publication (https://racs.uoregon.edu).

\bibliography{nucleation}

\end{document}